\documentclass[aps,prd,nofootinbib,superscriptaddress,12pt]{revtex4}
\usepackage{amsmath}
\usepackage{amssymb}
\usepackage{latexsym}
\usepackage{wasysym}
\usepackage{array}
\usepackage{graphicx}
\usepackage{color}

\newcommand{\eqrefer}[1]{Eq. \ref{#1}}
\newcommand{\figrefer}[1]{Fig. \ref{#1}}
\begin{document}
\bibliographystyle{hunsrt}

\date{\today}
\title{Probing Late Neutrino Mass Properties with Supernova
Neutrinos}
\author{Joseph Baker}
\affiliation{Department of Physics,
   University of Arizona,  Tucson AZ 85721}

\author{Haim Goldberg}
\affiliation{Department of Physics,
  Northeastern University,
Boston, MA 02115}

\author{Gilad Perez}
\affiliation{Theoretical Physics Group, Ernest Orlando Lawrence
Berkeley National Laboratory, University of California, Berkeley, CA
94720}
\affiliation{C.N. Yang Institute for Theoretical Physics
        State University of New York
        Stony Brook, NY 11794-3840
}

\author{Ina Sarcevic}
\affiliation{Department of Physics,
   University of Arizona,  Tucson AZ 85721}

\begin{abstract}
Models of late-time neutrino mass generation contain new
interactions of the cosmic background neutrinos with supernova relic
neutrinos (SRNs).   Exchange of an on-shell light scalar
may lead to significant modification of the differential SRN flux
observed at earth.  We consider an Abelian U(1) model for generating
neutrino masses at low scales, and  show that there are
cases for which the changes induced in the
flux allow one to
distinguish the Majorana or Dirac nature of neutrinos, as well as
the type of neutrino mass hierarchy (normal or inverted or
quasi-degenerate). In some region of parameter space the
determination of the absolute values of the neutrino masses is also
conceivable. Measurements of the presence of these effects may be
possible at the next-generation water Cerenkov detectors enriched
with Gadolinium, or a 100 kton liquid argon detector.
\end{abstract}

\maketitle

\section{Introduction}
Neutrino flavor conversion has been observed in the solar (SuperK,
SNO)~\cite{solar},
atmospheric (SuperK)~\cite{atmospheric}, and
terrestrial (KamLand,
K2K)~\cite{terrestrial}
neutrino data, providing evidence for
non-vanishing, sub-eV neutrino masses. There now remains the
longstanding theoretical question of how the neutrinos acquire their
masses. The most elegant solution to this puzzle is the seesaw
mechanism~\cite{seesaw}:
one assumes that lepton number is violated at some high scale
$\Lambda_{\rm L}$ in the form of right-handed neutrino, $N$, Majorana
masses, $M_N\sim \Lambda_{\rm L}$.  This induces, at a lower scale, an effective operator of
the form ${\cal O}(1)\times{(LH)^2/ \Lambda_{\rm L}}\,,$ where $L$
denotes a lepton doublet and $H$ the Higgs field. The oscillation
data then imply that $\Lambda_{\rm L}\sim 10^{14}\,{\rm GeV}\,$.
However, it is difficult to devise an experimental test of this
mechanism (see however~\cite{Buckley:2006nv}).
  Therefore, it is important to explore alternate
natural mechanisms for neutrino mass generation, especially those
that may be tested in experiments at low energies.

A class of such models that have astrophysical and cosmological
tests are the models of late-time neutrino mass
generation~\cite{Arkani-Hamed:1998pf,global,Davoudiasl:2005ks}. In
these models, neutrino masses are protected by some flavor symmetry
different from the one related to the charged fermion masses, for
example some global $U(1)_N$ symmetry. The small neutrino masses are
generated when the new symmetry is broken at low scales. The
effective Lagrangian for these models can be schematically written
for either Dirac or Majorana particles where the neutrino fields are
neutrino mass eigenstates as
\begin{eqnarray*}
{\cal L}_\nu^D = {\cal L}_{kin}+y_\nu \phi \nu N + V(\phi), \; \;
\; \; {\cal L}_\nu^M = {\cal L}_{kin}+y_\nu \phi \nu \nu +
V(\phi),
\end{eqnarray*}
where ${\cal L}_{kin}$ is the kinetic piece of the lagrangian,
$y_\nu$ is a dimensionless coupling, $\nu$ is a standard model
neutrino field, $N$ is an extra field introduced for the case of
neutrinos being Dirac particles, $\phi$ is the scalar field and
$V(\phi)$ is the associated scalar potential.  After spontaneous
symmetry breaking the neutrinos acquire masses given by
\begin{equation}
\label{neutrinomass}
m_\nu = y_\nu f \\
\end{equation} where $m_\nu$ is the mass of a
particular neutrino mass eigenstate and $f$ is the symmetry breaking
scale ($f$ = $\langle\phi\rangle$, where $\langle\phi\rangle$ is
the vacuum expectation value (VEV) of $\phi$). With just one
scalar the couplings are
diagonal in the mass basis.  In addition, since the scalar couples
to neutrinos only, the constraints on the symmetry breaking scale
$f$ are weak~\cite{global}.

In addition to generating neutrino mass through their VEVs,
the new light scalars provide another neutrino-neutrino interaction process
aside from the Standard Model $Z^0$ exchange. The effects of the
neutrinos coupling to these scalars on the cosmic microwave
background has been previously studied~\cite{global}. Constraints
have been placed on the symmetry breaking scale, $f$, the scalar mass,
$M_G$, and the scalar-neutrino couplings, $y_\nu$, in these models
from cosmological
considerations~\cite{global,Davoudiasl:2005ks,cosmologyconstraints},
as well as from demanding that supernova cooling and the flux of the
1987a neutrinos would not be significantly modified in the presence
of the additional fields~\cite{Farzan:2002wx,Goldberg:2005yw} (for
constraints related to generating the observed baryon asymmetry of
the universe see~\cite{Hall:2005aq}).

In this paper we show that the presence of this new physics
significantly modifies the spectrum of supernova relic neutrinos
(SRNs) at earth. This modification  occurs because SRNs  can
interact with cosmic background neutrinos through exchange of the
new light scalar. In~\cite{Goldberg:2005yw} this effect was studied
assuming a single flavor, Majorana, case. Here we extend our study
and include various interesting aspects  related to the nature of
the neutrino flavor sector, for example the  breaking of lepton
number, the effect of multiple generation, {\it etc.}  For
simplicity we confine our study here to late neutrino mass models
with a single $U(1)$ symmetry.  We show that the energy spectrum of
the SRN flux is sensitive to the type of neutrino mass hierarchy and
whether neutrinos are Majorana or Dirac particles.
 We discuss how in some
specific cases one can get additional information about the neutrino
masses as well.  In addition, detection of this signal would also be
direct evidence of the presence of the cosmic background neutrinos.
Hundreds of events per year from the flux of the SRN antineutrinos
could be seen at next-generation large megaton water Cerenkov
detectors~\cite{FogliBeacom,Hall:2006br} such as UNO,
Hyper-Kamiokande or MEMPHYS if they are enriched with
Gadolinium~\cite{nextgenexp}, or from the flux of the SRN neutrinos
in a large 100 kton liquid argon neutrino
detector~\cite{Ereditato:2005ru}.

In section \ref{SRN} we show how the SRN flux, including
cosmological evolution, is modified through the new interactions. In
section \ref{discussion} we consider the normal and inverted
neutrino mass hierarchy cases as well as quasi-degenerate neutrino
masses.  We also consider the possibility of neutrinos being either
Majorana or Dirac particles. In addition we show that there is a
particularly interesting signal that leads to the determination of
ratios of neutrino masses.  The conditions for establishing a
statistically significant signal above background are discussed in
subsection G of section III.  Finally, in section \ref{conclusion}, we
conclude.

\section{\label{SRN}The Supernova Relic Neutrino Flux}
The resonance interaction of the SRN with cosmic background
neutrinos through the exchange of a new light scalar was previously
discussed in~\cite{Goldberg:2005yw} for a single neutrino mass
eigenstate with $m_\nu \sim 0.05 eV$.

In this paper we consider the case of the scalar
interacting with three neutrino mass eigenstates and show
the very interesting effect on the observed
SRN spectrum for the normal mass hierarchy, inverted mass hierarchy,
and for quasi-degenerate neutrino masses. We briefly discuss the
case of adding one sterile neutrino.

We start with the SRN flux without the new interactions.  The
diffuse SRN flux is a remnant of neutrinos emitted from all the
supernova that have occurred in the universe~\cite{SRNstudies}. This
flux is given by
\begin{equation}
\label{SRNflux} F(E_\nu) = \int_0^{z_{max}} R_{SN}(z)
\frac{dN((1+z)E_\nu)}{dE_\nu}
\;(1+z)\left|c\frac{dt}{dz}\right| dz\;,\\
\end{equation}
where $R_{SN}(z)$ is the comoving rate of supernova formation,
$dN((1+z)E_\nu)/dE_\nu$ is the neutrino energy spectrum emitted by
supernova, $dt/dz$ is for the cosmological expansion, $c$ is the
speed of light, $z$ is the redshift, and $E_\nu$ is the neutrino
energy.

The quantity $R_{SN}$ is the comoving rate of supernova formation,
which can be parameterized as~\cite{SRNstudies}
\begin{equation}
\label{RSN}
R_{SN}(z)=\left(\frac{0.013}{M_{\odot}}\right)\stackrel{.}{\rho_*}(z),
\end{equation}
where $\stackrel{.}{\rho_*}(z)$ is the star formation rate given by
\begin{equation}
\label{rho} \stackrel{.}{\rho_*}(z) = (1-2) \times 10^{-2}M_\odot \;
yr^{-1} Mpc^{-3} \times (1+z)^\beta.
\end{equation}
We take ${\rm R_{{\rm SN}}}(0) = 2 \times 10^{-4} \, {\rm yr}^{-1}
\, {\rm Mpc}^{-3}$, $\beta = 2$ (for $0 < z <1$) and $\beta = 0$
(for $z
>1$)~\cite{SRNstudies}.  These are `median' values for the parameters which have uncertainties in
them coming from the uncertainty in the present knowledge of the
cosmic star formation
rate~\cite{SRNstudies,starformationrate}.\footnote{Future SN
observatories will have the power to significantly reduce the
related uncertainties~\cite{Hall:2006br}} The factor $dt/dz$ is
given by
\begin{equation}
\label{cosmology} \frac{dt}{dz}=-\left[100\frac{\rm km}{\rm
s~Mpc}\:h\:(1+z) \sqrt{\Omega_M(1+z)^3\; +
\Omega_\Lambda}\right]^{-1},
\end{equation}
with  $ \Omega_{M}=0.3 $, $ \Omega_\Lambda = 0.7 $ and $h$ = 0.7.

The energy spectrum of the neutrinos emitted by a supernova has been
modeled by several
groups~\cite{Totani:1997vj,Thompson:2002mw,Keil:2002in}. One of the
models leads to perfect equipartition of the energy radiated into each
neutrino flavor~\cite{Totani:1997vj}.  A second model makes a
detailed one dimensional calculation of all relevant neutrino
processes in the collapsing star and uses a variety of supernova
progenitor masses~\cite{Thompson:2002mw}.  Another model proposed by
Keil, Raffelt, and Janka~\cite{Keil:2002in} (KRJ model) performs
their calculations using a MC simulation.  These models give
spectra which have a width narrower than that of a thermal spectrum
(so-called ``pinched" spectrum~\cite{pinchedspectrum}). It also
predicts an average energy for the muon and tau flavor neutrinos
very close to the average energy of the electron antineutrinos. As
an illustrative example we take the neutrino spectrum given by the
KRJ model with the additional assumption that the total energy
carried by each neutrino flavor is $L_{\nu_e} = L_{\bar{\nu}_e} =
L_{\nu_x} = 5 \times 10^{52}$ ergs, where x stands for the muon and
tau neutrinos and antineutrinos. The KRJ energy spectrum of neutrino
flavor eigenstates produced by a supernova is given by
\begin{equation} \label{dNdE} \frac{dN_{\nu_\alpha}(E_{\nu_\alpha})}{dE_{\nu_\alpha}} =
\frac{(1+\beta_{\nu_\alpha})^{1+\beta_{\nu_\alpha}}L_{\nu_\alpha}}{\Gamma[1+\beta_{\nu_\alpha}]\overline{E_{\nu_\alpha}}^2}
\left( \frac{E_{\nu_\alpha}}{\overline{E_{\nu_\alpha}}}
\right)^{\beta_{\nu_\alpha}}
\exp{[-(1+\beta_{\nu_\alpha})\frac{E_{\nu_\alpha}}{\overline{E_{\nu_\alpha}}}],}\\
\end{equation}
with the $\overline{E_{\nu_\alpha}}$ representing the average
neutrino energies and $\beta_{\nu_\alpha}$ characterizing the
amount of spectral pinching.  Here for simplicity we set the values
of the parameters to~\cite{Keil:2002in} (for recent numerical
studies see {\it e.g.}~\cite{numericalstudies})
\begin{eqnarray}
\label{KRJparams}
\nu_e: \; \beta_{\nu_e} = 3.4 \; \; \; \overline{E_{\nu_e}} = 13.0 \; \rm MeV, \nonumber \\
\bar{\nu}_e: \; \beta_{\bar{\nu}_e} = 4.2 \; \; \; \overline{E_{\bar{\nu}_e}} = 15.4 \; \rm MeV, \\
\nu_x: \; \beta_{\nu_x} = 2.5 \; \; \; \overline{E_{\nu_x}} = 15.7
\; \rm MeV, \nonumber
\end{eqnarray}
and also neglect effects such as shock wave and
turbulence~\cite{turbulence}.

Because of matter oscillation effects, neutrinos emerge from a
supernova as coherent fluxes of mass eigenstates which we label as
$F_{\nu_i}$, where $i$ = 1, 2, or 3  represents the particular
neutrino mass eigenstate~\cite{Dighe:1999bi}.

If neutrino flavor evolution inside of the collapsing star is
either fully adiabatic or fully non-adiabatic (the flavor evolution
is adiabatic if the mixing angle $sin^2\theta_{13} \gtrsim 10^{-3}$
and non-adiabatic if $sin^2\theta_{13}\lesssim 10^{-5}$) then the
energy spectrum of each neutrino mass eigenstate that leaves the
surface of the star corresponds to the original energy spectrum of
some particular neutrino flavor eigenstate at emission from the
neutrinosphere, {\em i.e.,} there is a one-to-one correspondence
between each $dN_{\nu_\alpha}(E_{\nu_\alpha}) /dE_{\nu_\alpha}$ and
some $dN_{\nu_i}(E_{\nu_i})/dE_{\nu_i}$.\footnote{ See, for example,
Table 1, Fogli, et. al. \cite{nudecay}}
The original produced
flux of some neutrino flavor at the neutrinosphere will be
labeled as $F_{\nu_\alpha}^0$.  Translated back into the flavor basis,
the expressions for the
$\nu_e$ and $\bar{\nu}_e$ fluxes emerging from a supernova can be
written as
\begin{eqnarray}\label{normalfluxes}
F_{\nu_e}=P_H|U_{e2}|^2F^0_{\nu_e}+(1-P_H|U_{e2}|^2)F^0_{\nu_x}, \\
F_{\bar{\nu}_e}=|U_{e1}|^2F^0_{\bar{\nu}_e}+|U_{e2}|^2F^0_{\nu_x}, \; \; \;
\; \; \; \; \; \; \; \; \; \; \; \; \; \; \; \; \nonumber
\end{eqnarray}
for the normal mass hierarchy and
\begin{eqnarray}\label{invertedfluxes}
F_{\nu_e}=|U_{e2}|^2F^0_{\nu_e}+|U_{e1}|^2F^0_{\nu_x}, \; \; \; \; \;
\; \;
\; \; \; \; \; \; \; \; \; \; \; \\
F_{\bar{\nu}_e}=\bar{P}_H|U_{e1}|^2F^0_{\bar{\nu}_e}+(1-\bar{P}_H|U_{e1}|^2)F^0_{\nu_x},\nonumber
\end{eqnarray}
for the inverted mass hierarchy, where $P_H({\rm and} \; \bar{P}_H)
= 0$ for the adiabatic case and $P_H({\rm and} \; \bar{P}_H) = 1$
for the non-adiabatic case~\cite{Dighe:1999bi}. In the equations
above, $|U_{e1}|^2 = \cos^2{\theta_{12}},\; |U_{e2}|^2 =
\sin^2{\theta_{12}},\; \theta_{12} = \theta_{\tiny\astrosun}$
($\theta_{\tiny\astrosun}$ is the solar mixing
angle~\cite{GGarcia}), where $\sin^2{2 \theta_{12}} = 0.86
\pm 0.3$, and $|U_{e3}|^2 \approx 0$~\cite{PDBook}. When supernova
neutrino flavor evolution is non-adiabatic then the $\nu_e$ and
$\bar{\nu}_e$ flux for the normal and inverted hierarchies are
identical.

We show that  the addition of  a new light
scalar opens the possibility of determining the neutrino mass
hierarchy independent of the neutrino fluxes, and also independent
of the adiabatic or non-adiabatic nature of supernova neutrino
flavor evolution.

\subsection{\label{modified}Modifications due to New Physics}
We consider the modifications to the SRN flux due to the resonance
interaction of an SRN with a neutrino in the cosmic neutrino
background. In this process a supernova neutrino with energy
$E_\nu^{SN}$ will go through the resonance when the kinematic
condition
\begin{equation}
\label{rescondition} E_{\nu}^{SN} = \frac{M_G^2}{2 m_i} \equiv
E_i^{Res},
\end{equation}
is satisfied.  More specifically, a neutrino observed with energy
$E_\nu^{Obs}$ will have gone through resonance if its energy lies
in the region
\begin{equation}
\label{reswindow} \frac{E_i^{Res}}{1+z} < E_\nu^{Obs} < E_i^{Res},
\end{equation}
where $z$ is the redshift.  There will be large depletion of the SRN
flux in the energy domain given in \eqrefer{reswindow} as long as
the neutrino-scalar coupling satisfies~\cite{Goldberg:2005yw}
\begin{equation}
\label{lowercoupling} y > 4.6 \times 10^{-8} \frac{M_G}{1\; \rm
keV}\;.
\end{equation} This condition comes from requiring that the mean free path for
absorption is much smaller than the Hubble scale.
It is important to note that in the narrow width approximation
for the resonance, the condition \eqrefer{lowercoupling} is a sufficient
condition to guarantee the absorption of all three neutrino flavors.
After the neutrinos that have energies in the region given by
\eqrefer{reswindow} go through the resonance they are
redistributed to lower energies when the produced scalar decays
back to neutrino mass eigenstates~\cite{instdecay}. In particular, neutrinos after
interaction will be redistributed with a flat energy distribution
from zero energy up to the original energy of the incident
supernova neutrino.

To find the effect of these interactions on the flux of the SRN we
note that the neutrinos leaving a supernova at redshift $z$ emerge
as  the mass eigenstates. However, these mass eigenstate fluxes are
now modified through interaction with the cosmic background
neutrinos as they propagate to the earth. We consider for simplicity
an Abelian U(1) late neutrino mass model.  This implies that the
Yukawa interaction between the scalars (in particular the Goldstone)
and the neutrinos are diagonal in the mass basis (this is not the
case in a model with non-Abelian symmetries).  Supernova neutrinos
are in their mass eigenstates and each mass eigenstate interacts
only with the same mass eigenstate background neutrino via Goldstone
exchange. To illustrate how the interactions modify the neutrino
mass eigenstate flux we consider as an example the flux of the
$\nu_1$ mass eigenstate, $F_{\nu_1}.$

We start by defining the modified flux of the $\nu_1$ eigenstates as
$\widetilde{F_{\nu_1}}$.  For each redshift $z$ the $\nu_1$ eigenstates
that satisfy the condition given by \eqrefer{reswindow} will have
resonance interaction with cosmic background neutrinos, producing
the intermediate scalar. A neutrino mass eigenstate will go
through the resonance when the coupling satisfies
\eqrefer{lowercoupling}. The cross section (averaged over
the width of the resonance) for this to
occur is approximately given by $\sigma_{Res}\simeq\pi/M^2_G$.  This
will lead to a mean free path much smaller than the typical distance a
supernova neutrino will travel to arrive at the earth, for the values
of $M_G$ that we consider.

We label the absorbed flux as $F_{\nu_1}^{Res}$.
Naively, the modified flux would be given by
\begin{equation}
\label{resstep1} \widetilde{F_{\nu_1}} = F_{\nu_1} -
F_{\nu_1}^{Res}.
\end{equation}
However, this expression does not take into account that the scalar
decays back into neutrino mass eigenstates.  We need to add this
contribution to \eqrefer{resstep1}. The scalar can decay to any of
the neutrino mass eigenstates. The probability that the scalar
decays to a particular neutrino mass eigenstate is proportional to
the square of the Yukawa coupling of that particular neutrino mass
eigenstate to the scalar. From \eqrefer{neutrinomass} we note that
the relative probabilities are proportional to the ratios of squares
of the neutrino masses,
\begin{equation}
\label{probability} P_j \approx \frac{m_j^2}{\sum_{i=1}^3 m_i^2}.
\end{equation}
Then the probability that a scalar decays to the neutrino mass
eigenstate $\nu_1$ is $P_1$.  These decays result in redistribution
of the neutrino energies from zero energy up to the energy of the
incident SRN with a flat energy distribution.  We define $P_1 \times
F_{1\rightarrow 1'}^{Res}$ as the fraction of the flux of $\nu_1$
that initiate a resonance, producing a scalar which then decays back
into a $\nu_1$ eigenstate with degraded energy (indicated by the
notation $1'$).  Then, \eqrefer{resstep1} is modified to
\begin{equation}
\label{resstep2} \widetilde{F_{\nu_1}} = F_{\nu_1} - F_{\nu_1}^{Res}
+ P_1 \times F_{1\rightarrow 1'}^{Res}.
\end{equation}
We still need to take into account the contributions from the decays
of scalars produced by other neutrino mass eigenstates. Therefore,
there should be a sum over all of the initial states, and
\eqrefer{resstep2} becomes
\begin{equation}
\label{resstepfinal} \widetilde{F_{\nu_1}} = F_{\nu_1} -
F_{\nu_1}^{Res} + P_1 \sum_{i=1,2,3,\bar{1},\bar{2},\bar{3}}
F_{i\rightarrow 1'}^{Res}.
\end{equation}
In more general notation, for the $j^{th}$ neutrino mass
eigenstate,
\begin{equation}
\label{modifiedmassflux} \widetilde{F_j}=F_j-F_j^{res}+ P_j \times
\sum_{i=1,2,3,\bar{1},\bar{2},\bar{3}} F_{i \rightarrow j'}^{Res} \;
\;.
\end{equation}
The contributions over a range of redshift must be taken to
determine the total flux at earth.  If neutrinos are Dirac particles
then there is factor of 1/2 multiplying the last term (see
discussion in Section \ref{diracVSmajorana}).

The modified flux of electron neutrinos and electron antineutrinos
can then be written as
\begin{equation}
\label{modifiednueflux} \widetilde{F_{\nu_e}} = \cos^2{\theta_{12}}
\widetilde{F_{\nu_1}} + \sin^2{\theta_{12}}\widetilde{F_{\nu_2}},
\end{equation}
and
\begin{equation}
\label{modifiednuebarflux} \widetilde{F_{\bar{\nu}_e}} =
\cos^2{\theta_{12}} \widetilde{{F}_{\bar{\nu}_1}} +
\sin^2{\theta_{12}}\widetilde{{F}_{\bar{\nu}_2}}.
\end{equation}
Finally, we note that each neutrino mass eigenstate goes through
resonance at different energies given by \eqrefer{rescondition} when
there is just a single scalar of mass $M_G$.  Depending on the
details of the neutrino mass hierarchy, these resonance energies can
either be very close to one another, or widely spaced apart.

\section{\label{discussion}Signals of Models of Late-Time Neutrino Mass Generation}
In this section we discuss the signals for the neutrino mass
hierarchy in the observed SRN flux.  We consider the case of the
normal neutrino mass hierarchy, the inverted neutrino mass
hierarchy, and the possibility that the neutrino masses are
quasi-degenerate. We also show the effects of the neutrinos being
Dirac or Majorana particles on the SRN flux signal.  For these
cases, unless otherwise noted, we choose the value of $E_i^{Res}$,
defined in \eqrefer{rescondition}, at $z=0$ to be equal to 15 MeV
for one of the neutrino mass eigenstates. This choice is made to
illustrate the effects of the resonance process and to determine a
region of the parameter space of the late-time neutrino mass
generation models where the effect of the SRN modification would be
seen.  This is so, since for water Cerenkov detectors near reactors the
background becomes negligible above about 13 MeV~\cite{FogliBeacom,Hall:2006br}.
As will be discussed in the conclusions, the energy resolution is about
2 MeV.  In subsections A through E we consider the case
where neutrino flavor evolution in the SN is adiabatic, and in subsection F we
show that the same features are obtained for a case where the
flavor evolution is nonadiabatic.  In subsection G we discuss the
detection of the new interactions.

In the following we focus on the flux of electron antineutrinos that
arrive at earth, since the proposed water Cerenkov
experiments for detection are
sensitive to this neutrino flavor through the interaction of
electron antineutrinos with protons with a cross-section given
by~\cite{CrosssectionRefs}
\begin{equation}
\label{crosssection} \sigma = 10^{-43} p_{e^+}E_{e^+}
E_\nu^{-0.07056+0.02018\ln{E_\nu}-0.001953\ln^2{E_\nu}} {\rm cm}^2,
\end{equation}
where $E_{e^+}$ and $p_{e^+}$ are the energy and momentum of the
detected positron.  Note that detection of the electron neutrino
component of the SRN flux at a large liquid argon detector
would provide complementary information~\cite{SRNnue}.

\subsection{\label{Normal}Normal Neutrino Mass Hierarchy}
\begin{figure}[h]
\begin{center}
\includegraphics[width=4in]{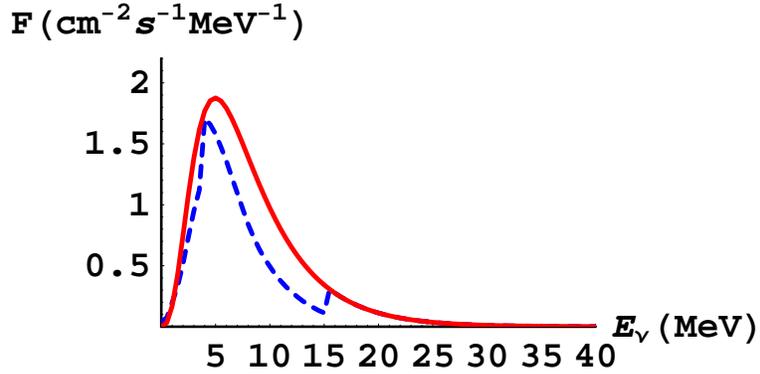}
\caption{The SRN electron antineutrino flux without interactions
(red (solid) curve) and with interactions (blue (dashed) curve) when neutrinos are
Majorana particles and for the normal mass hierarchy.}
\label{majnormalM1}
\end{center}
\end{figure}

\begin{figure}[h]
\begin{center}
\includegraphics[width=4in]{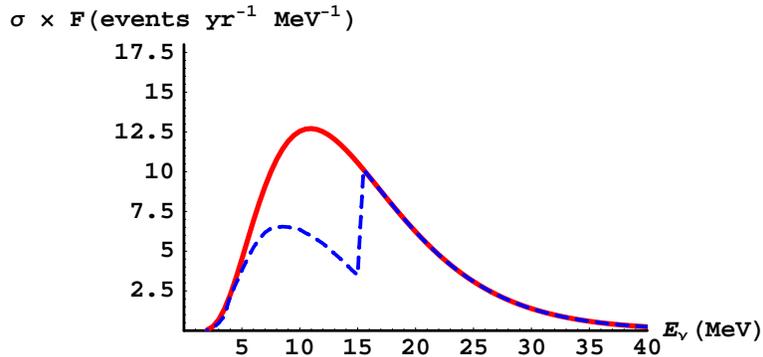}
\caption{The event rates for HyperKamiokande without interactions
(red (solid) curve) and with interactions (blue (dashed) curve) when neutrinos are
Majorana particles and for the normal mass hierarchy.}
\label{majnormalM1weighted}
\end{center}
\end{figure}

As an example to illustrate how resonance interactions between the
SRN and the cosmic background neutrinos can affect the SRN flux, we
first consider a normal mass hierarchy of neutrino mass eigenstates.
As a particular example of this hierarchy we choose the masses of
the mass eigenstates to be
\begin{eqnarray}
\label{normalhierarchy}
m_1 = 0.002 \; \rm eV, \nonumber \\
m_2 = 0.009 \; \rm eV, \\
m_3 = 0.05 \;\;\; \rm eV. \nonumber
\end{eqnarray}
This conforms to the best value of the atmospheric mass splitting
$\vert m_3^2-m_{1,2}^2\vert \simeq 2.4 \times 10^{-3} \rm
eV^2$~\cite{Fogli:2005gs}.
The value of the lightest mass, $m_1$, was chosen to be 0.002 eV for
the purpose of our numerical study, however there is no lower limit
on the value of the mass of the lightest neutrino in either the
normal or inverted hierarchies.  If the mass of the lightest
neutrino is lowered below the neutrino background temperature,
$T_{C\nu B}$, then the resonance for this lightest state is governed by the
corresponding thermal energy of the background neutrinos (see footnote 2
below).
The particular choice for the
neutrino masses in \eqrefer{normalhierarchy} results in the following features:
\begin{enumerate}
    \item When a scalar is produced, it decays predominantly to
    the $m_3$ mass eigenstate.  \eqrefer{probability} gives $P_3
    \approx 0.967, \; P_2 \approx 0.031,$ and $P_1 \approx 0.002$.
    \item Because there is one scalar with mass $M_G$,
    \eqrefer{rescondition} implies that the ratios of the neutrino
    masses govern the resonance energy positions, so that
    $E_2^{Res} = 2/9 \times E_1^{Res}$ and $E_3^{Res} = 1/25
    \times E_1^{Res}$.\footnote{{ The value of $E_1^{Res}$ for the
lightest neutrino could be in the range
$\sqrt{2 E_\nu T_{C\nu B}} \lesssim E_1^{Res} \leq \sqrt{2 E_\nu
m_1}$, where the lower limit corresponds to the transition to
the relativistic case and $T_{C\nu B}$ is the background neutrinos
temperature. Since $T_{C\nu B}\sim 2\times10^{-4}$ which is not far from
$m_1$ (given that the effect goes like the square root of the mass in
that range) our results will only be slightly modified when this is taken
into account.}}
\end{enumerate}
If we choose the lightest neutrino mass eigenstate to have the
resonance at 15 MeV, to illustrate the effect, then the
corresponding scalar mass is $M_G \approx 245$ eV, and point 2
above implies that the other two resonances are
$E_2^{Res} \approx 3.8$ MeV and $E_3^{Res} \approx 0.60$ MeV. These
two resonance energies are both well below experimental detection thresholds.
Additionally, because the modified electron antineutrino flux is
composed only of $\widetilde{F_{\bar{\nu}_1}}$ and
$\widetilde{F_{\bar{\nu}_2}}$ the overall effect on
$\widetilde{F_{\bar{\nu}_e}}$ is a depletion since decays of the
scalar primarily contribute to $\widetilde{F_{\bar{\nu}_3}}$.  The
resulting flux of electron antineutrinos can be seen in
\figrefer{majnormalM1}.  Folding the flux with the cross-section for
electron antineutrinos on protons,\eqrefer{crosssection}, gives the
spectrum in \figrefer{majnormalM1weighted}.


Relative to the electron antineutrino flux without interactions, the
case with interactions has large depletion because of the dominant
decay into the $m_3$ mass eigenstate. If we choose a heavier scalar
so that the $m_2$ mass eigenstate goes through the resonance at 15
MeV instead of the $m_1$ mass eigenstate (i.e., $M_G \approx 490$
eV), then we get the same feature, but the depletion is smaller.
This is because $\widetilde{F_{\bar{\nu}_1}}$ is multiplied by
$\cos^2{\theta_{12}} \approx 0.70$, while
$\widetilde{F_{\bar{\nu}_2}}$ is multiplied by $\sin^2{\theta_{12}}
\approx 0.30$, so $\widetilde{F_{\bar{\nu}_2}}$ is the smaller
component of the final electron antineutrino flux. If the scalar
were even heavier (i.e., $M_G \approx 1.2$ keV) so that the $m_3$
mass eigenstate goes through the resonance at 15 MeV, then there is
no depletion in the electron antineutrino flux since both the $m_1$
and $m_2$ eigenstates would have resonances at very high energies.
As a result, neither of the corresponding fluxes would be visibly
modified.

The mass of the lightest neutrino can also be lowered without
bound (although as mentioned above we do not consider the cases where the mass of the
lightest neutrino is below the mass of the cosmic background
neutrino temperature).  For a fixed value of $M_G$, as the mass of
the lightest state is lowered, the position of the resonance moves
to higher energies until it is in a region where the flux of the SRN
is too small for the feature to be experimentally observable.

\begin{figure}[h]
\begin{center}
\includegraphics[width=4in]{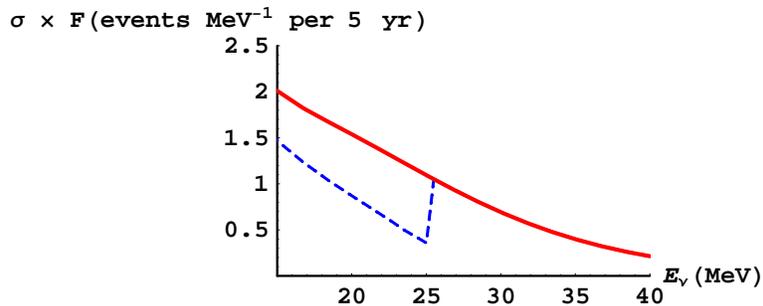}
\caption{The SRN electron neutrino flux for a normal hierarchy folded
with the cross-section for electron neutrinos to interact with argon
in a 100 kton detector running for 5 years.
The red (solid) curve is for no interactions and the blue (dashed)
curve is with interactions.}
\label{liquidargon}
\end{center}
\end{figure}

To illustrate the effect on the electron neutrino spectrum, relevant
to argon detectors such as Icarus and a future 100 kton argon detector~\cite{Ereditato:2005ru},
we show the event rates in \figrefer{liquidargon} for the normal hierarchy case.  A next generation liquid argon
detector with a size of 100 ktons could measure a significant number of
electron neutrinos over just 5 years~\cite{Ereditato:2005ru}.  At neutrino energies lower
than about 19 MeV the solar neutrino flux dominates the SRN flux, and at
energies greater than about 40 MeV the atmospheric neutrino flux begins to
dominate.  In \figrefer{liquidargon}, to obtain event rates, we have folded the SRN flux
with the cross-section for electron neutrinos to interact with argon~\cite{Ereditato:2005ru}.
We have used a resonance energy of 25 MeV for the
lightest neutrino mass state (taken to be 0.001 eV so that $M_G \approx 225 \, {\rm eV}$).
In this case we find significant reduction in the integrated event
rate over the region in which the SRN flux is dominant.

\subsection{\label{Inverted}Inverted Neutrino Mass Hierarchy}
\begin{figure}[h]
\begin{center}
\includegraphics[width=4in]{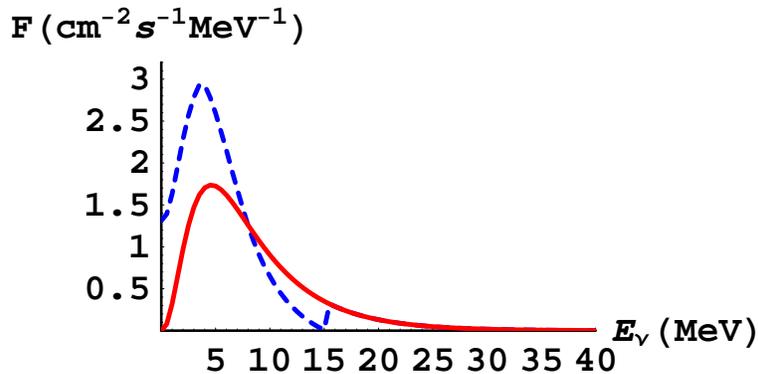}
\caption{The SRN electron antineutrino flux without interactions
(red (solid) curve) and with interactions (blue (dashed) curve) when neutrinos are
Majorana particles and for the inverted mass hierarchy.}
\label{majinvertedM12}
\end{center}
\end{figure}

\begin{figure}[h]
\begin{center}
\includegraphics[width=4in]{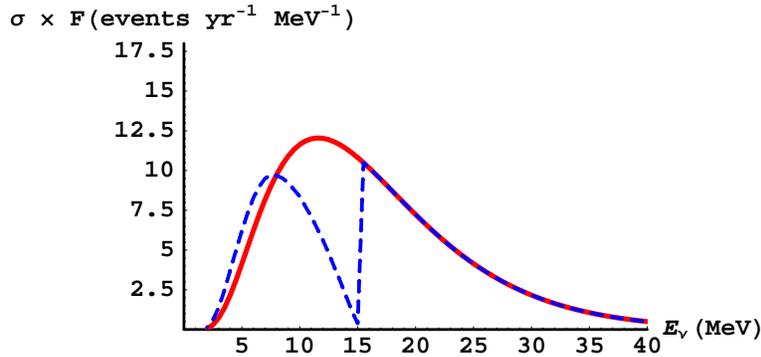}
\caption{The event rates for HyperKamiokande without interactions
(red (solid) curve) and with interactions (blue (dashed) curve) when neutrinos are
Majorana particles and for the inverted mass hierarchy.}
\label{majinvertedM12weighted}
\end{center}
\end{figure}

We now consider an example of the inverted mass hierarchy, characterized
by  $m_1\simeq m_2 >> m_3$,  with
neutrino masses chosen to be
\begin{eqnarray}
\label{invertedhierarchy}
m_1 = 0.05 \; \rm eV, \; \;\nonumber \\
m_2 \approx 0.05 \; \rm eV, \; \; \\
m_3 = 0.008 \;  \rm eV. \nonumber
\end{eqnarray}
This reflects the best value of $7.92 \times 10^{-5} \; \rm
eV^2$~\cite{Fogli:2005gs} for $\Delta m_{21}^2,$ the $\nu_1$-$\nu_2$
mass splitting. Whenever a scalar is produced  it
dominantly decays into these two heavy eigenstates with equal
probabilities. Because $\widetilde{F_{\bar{\nu}_1}}$ and
$\widetilde{F_{\bar{\nu}_2}}$ are the contributing components to the
electron antineutrino flux, in this scenario there are both regions
of depletion but also regions of overall enhancement of the flux due to
rescattering. We
choose the case where the $m_1$ and $m_2$ mass eigenstates go
through resonance at 15 MeV, giving $M_G \approx 1.2 \; \rm keV$, and
show the results in \figrefer{majinvertedM12} (and
\figrefer{majinvertedM12weighted} for weighting with cross-section).


We find that the enhancement is large because all initial neutrino
mass eigenstate fluxes produce scalars which add to the low energy
$m_1$ and $m_2$ eigenstate fluxes.  The $m_1$ and $m_2$ eigenstate
flux is depleted and is redistributed to lower energies. Once we
fold the flux with the cross-section, we see in
\figrefer{majinvertedM12weighted} that in contrast to the case of
the normal mass hierarchy, the peak at low energies is more
pronounced.  This is a result of the electron antineutrino flux
being composed only of the $m_1$ and $m_2$ neutrino mass eigenstates
(since $|U_{e3}|^2 \approx 0$), both of which get depleted by the
resonance at the same energy since the two eigenstates are nearly
mass degenerate. Therefore the electron antineutrino flux is almost
completely depleted near the resonance cutoff.

\begin{figure}[h]
\begin{center}
\includegraphics[width=4in]{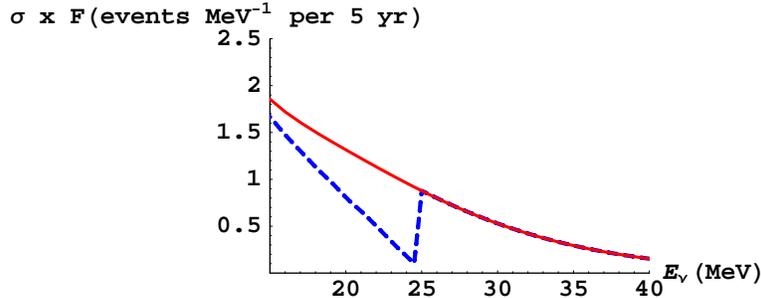}
\caption{The SRN electron neutrino flux for an inverted hierarchy folded
with the cross-section for electron neutrinos to interact with argon
in a 100 kton detector running for 5 years.
The red (solid) curve is for no interactions and the blue (dashed)
curve is with interactions.}
\label{liquidargoninverted}
\end{center}
\end{figure}

We apply similar analysis to the electron neutrino flux, of relevance
to a liquid argon detector.
In \figrefer{liquidargoninverted} we show event rates for 100kton
liquid argon detector.
We use a resonance energy of 25 MeV
for the two heavier mass eigenstates with mass 0.05 eV, which corresponds to
 $M_G \approx 1580 \, {\rm eV}$.  The
 integrated event rate for energies above the solar background are reduced
relative to the no interaction case.

\subsection{\label{Degenerate}Quasi-Degenerate Neutrino Masses}
\begin{figure}[h]
\begin{center}
\includegraphics[width=4in]{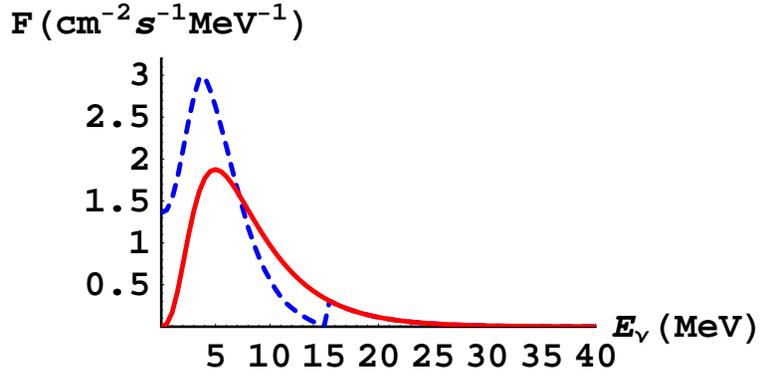}
\caption{The SRN electron antineutrino flux without interactions
(red (solid) curve) and with interactions (blue (dashed) curve) when neutrinos are
Majorana particles and for quasi-degenerate neutrino
masses.}
\label{degenerate}
\end{center}
\end{figure}

\begin{figure}[h]
\begin{center}
\includegraphics[width=4in]{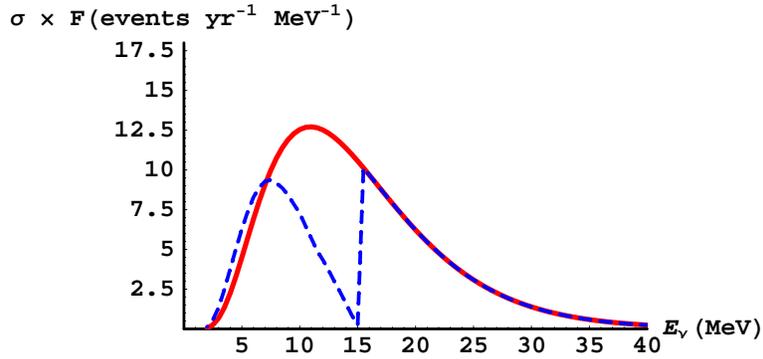}
\caption{The event rates for HyperKamiokande without interactions
(red (solid) curve) and with interactions (blue (dashed) curve) when neutrinos are
Majorana particles and for quasi-degenerate neutrino masses.}
\label{degenerateweighted}
\end{center}
\end{figure}

There still remains the possibility that the neutrino mass
eigenstates are quasi-degenerate.  For example, a mass hierarchy
structure with
\begin{eqnarray}
\label{degeneratehierarchy}
m_1 \approx 0.06 \; \rm eV, \nonumber \\
m_2 \approx 0.06 \; \rm eV, \\
m_3 \approx 0.08 \; \rm eV, \nonumber
\end{eqnarray}
satisfies the requirements for the two independent mass splittings
as well as cosmological constraints on the sum of the neutrino
masses~\cite{Goobar:2006xz}. We consider the case where the two
eigenstates with mass 0.06 eV have the same resonance energy as before,
which would correspond to $M_G = 1340 \; \rm eV$. The third mass
eigenstate then has a resonance energy at approximately 11 MeV,
however there is no corresponding dip since only the $m_1$ and $m_2$ mass
eigenstates contribute to the final $\bar{\nu}_e$ flux. The result for the
flux can be seen in \figrefer{degenerate} and for the event rate in
\figrefer{degenerateweighted}.


It is clear by comparing \figrefer{majinvertedM12} and
\figrefer{degenerate} that the case of the inverted mass hierarchy
is nearly indistinguishable from the case of quasi-degenerate
neutrino masses.  However the quasi-degenerate case is
distinguishable from the specific case of the normal mass hierarchy
when one neutrino mass is much lighter than the other two masses.

\subsection{\label{multiplefeatures}Multiple Depletion Dips}
There is the possibility within the normal hierarchy that the $m_1$
and $m_2$ mass eigenstates are nearly, but not exactly, degenerate,
and also still much lighter than the $m_3$ mass eigenstate.  One
example of the possible values for the masses in such a scenario is
\begin{eqnarray}
\label{multiplehierarchy}
m_1 \approx 0.01 \; \rm eV, \; \; \nonumber \\
m_2 \approx 0.013 \; \rm eV, \\
m_3 \approx 0.05 \; \rm eV. \; \; \nonumber
\end{eqnarray}
If we choose the resonance of the $m_2$ eigenstate to be at 15 MeV,
then the $m_1$ mass eigenstate goes through resonance at $E_1^{Res}
\approx 20 \; \rm MeV$.  This corresponds to $M_G \approx 630 \; \rm
eV$.

\begin{figure}[h]
\begin{center}
\includegraphics[width=4in]{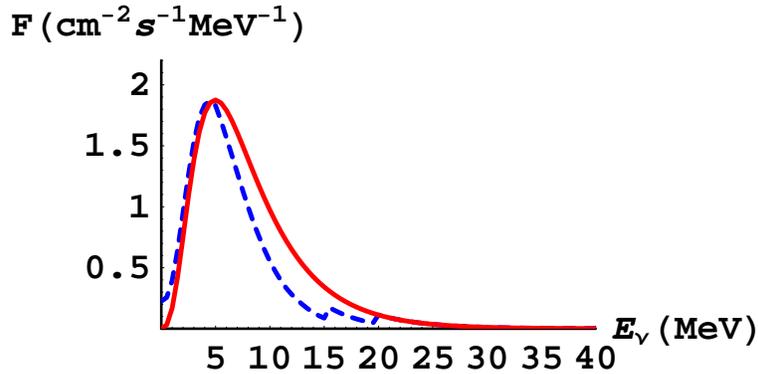}
\caption{The SRN electron antineutrino flux without interactions
(red (solid) curve) and with interactions (blue (dashed) curve) when neutrinos are
Majorana particles, for the normal mass hierarchy and where two
neutrinos have distinct resonance features in the experimentally
observable region.} \label{doubleres}
\end{center}
\end{figure}

\begin{figure}[h]
\begin{center}
\includegraphics[width=4in]{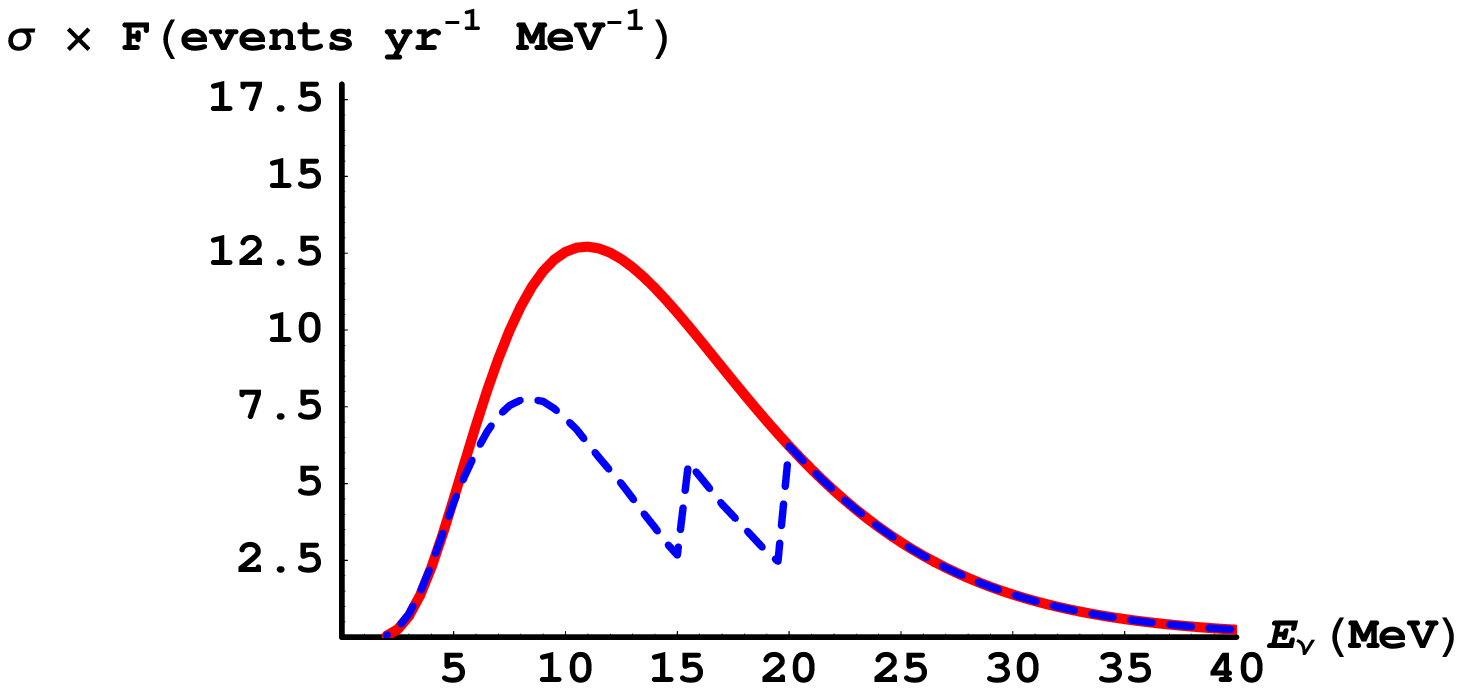}
\caption{The event rates for HyperKamiokande without interactions
(red (solid) curve) and with interactions (blue (dashed) curve) when neutrinos are
Majorana particles, for the normal mass hierarchy and where two
neutrinos have distinct resonance features in the experimentally
observable region.}\label{doubleresweighted}
\end{center}
\end{figure}

As can be seen in \figrefer{doubleres} and
\figrefer{doubleresweighted}, this leads to two depletion dips in the
final electron antineutrino spectrum and three peaks.  There is
always a signal corresponding to each neutrino mass eigenstate
interacting with the Goldstone.  The presence of two distinct
depletion dips in an experimentally
interesting region, however, is sensitive to the ratio of the masses
of two of the neutrino mass eigenstates, in this case $m_1$ and
$m_2$.  For example, in Section \ref{Normal} there is only one
depletion dip because the ratio of the masses, and therefore the
ratio of the resonance energies, for the $m_1$ and $m_2$ states is
4.5, so that with $m_1$ resonance at 15 MeV the $m_2$ resonance is
at 3.8 MeV, outside of the observable region.

Experimental observation of the energy position of these
dips could determine the ratio of the $m_1$ and
$m_2$ masses, which together with the measured value of $\Delta
m_{21}^2$ and $\Delta m_{32}^2$ allows one to determine the neutrino
masses.  \noindent This is a remarkable
possibility since it is extremely hard to experimentally determine
the exact values of the neutrino masses, especially the mass of the
lightest state.


\subsection{\label{diracVSmajorana}Dirac vs. Majorana Neutrinos}
If neutrinos are Majorana particles, then each scalar decay produces
a $\nu_L \nu_L$ or $\nu_R \nu_R$ for each mass eigenstate. If the
neutrinos are Dirac particles then the scalar can decay to
$\nu\bar{N}$ or to ${N}\bar{\nu}$, where $\bar{N}$ and $N$ are the
extra neutrino fields added for the case of neutrinos being Dirac
particles in the late-time neutrino mass generation models.
Then only half of the decays of the scalar produce an antineutrino
that will be seen in the detector. Therefore for the case of
neutrinos being Dirac particles there is an overall factor of $1/2$
multiplying the last term of \eqrefer{modifiedmassflux} relative to
the case of neutrinos being Majorana particles.

\begin{figure}[h]
\begin{center}
\includegraphics[width=4in]{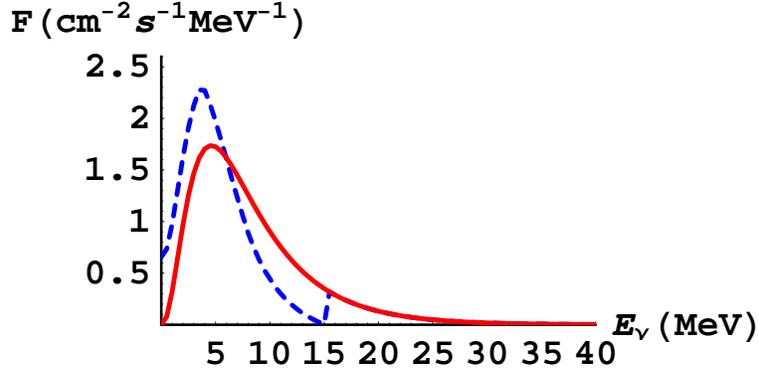}
\caption{The SRN electron antineutrino flux without interactions
(red (solid) curve) and with interactions (blue (dashed) curve) when neutrinos are
Dirac particles and for the inverted mass hierarchy.}
\label{diracinvertedM12}
\end{center}
\end{figure}

If the neutrinos are arranged in a normal mass hierarchy as in
Section \ref{Normal} then the ability to distinguish between the
neutrinos being Dirac or Majorana particles is confounded by the
small amount of scalar decays into the $m_1$ and $m_2$ eigenstates.
However, in the case of the inverted mass hierarchy of Section
\ref{Inverted} there can be a visible difference in the electron
antineutrino flux if the neutrinos are Dirac or Majorana particles
as seen by comparing the Majorana particle case of
\figrefer{majinvertedM12} with the Dirac particle case of
\figrefer{diracinvertedM12}.

The resonance energy of the $m_1$ and $m_2$ mass eigenstates have
been set to 15 MeV in this case, exactly the same as for the
Majorana particle case considered in Section \ref{Inverted}. Because
of the extra factor of $1/2$ the overall scale of the enhancement is
much smaller than in the case of neutrinos being Majorana particles.

\subsection{\label{nonadiabatic}Non-adiabatic case ($sin^2\theta_{13}\lesssim 10^{-5}$)}
If supernova neutrino flavor evolution is non-adiabatic, then
the flux of electron antineutrinos that leaves a supernova is
independent of the neutrino mass hierarchy (see \eqrefer{normalfluxes} and
\eqrefer{invertedfluxes}). If the
supernova neutrinos interact with the cosmic background neutrinos
via new light scalars then the flux observed at earth will be
different for the normal mass hierarchy and the inverted mass
hierarchy.  However, this difference is not detectable at the present
or near-future neutrino experiments because it relies on the ability
to observe neutrinos in the SRN flux at low energies where there is a large
reactor background.

The difference between the two neutrino mass hierarchies is present,
even in the case of small $sin^2\theta_{13}$, because for a normal mass hierarchy,
the heavy neutrino mass eigenstate is $m_3$ which does not contribute to the $\nu_{\bar{e}}$
flux.  All of the scalars produced through the neutrino-neutrino
interactions will dominantly decay into this heaviest neutrino mass
eigenstate, and the final flux will have an overall depletion
relative to the SRN flux without interactions.  However, for the
inverted hierarchy, the $m_1$ and $m_2$ mass
eigenstates are the heavy states, while the $m_3$ mass eigenstate is
the light state.  The scalars produced through the neutrino-neutrino
interactions will dominantly decay into the two heavy states,
leading to a low energy enhancement as well as the higher energy
dip.

\begin{figure}[h]
\begin{center}
\includegraphics[width=4in]{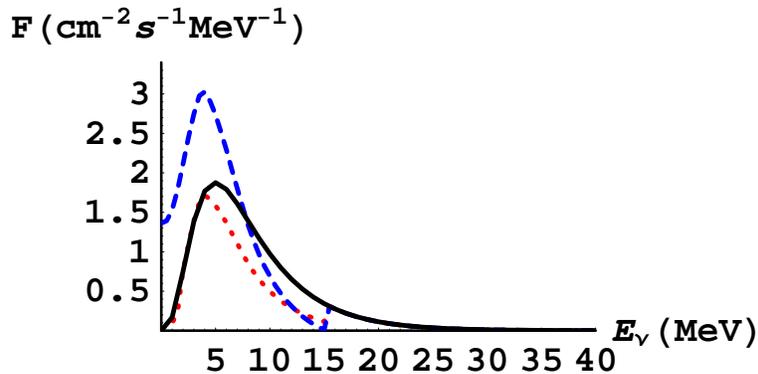}
\caption{The SRN electron antineutrino flux without interactions
(black solid curve), with interactions and normal mass hierarchy
(red dotted curve), and with interactions and inverted mass
hierarchy (blue dashed curve), when neutrinos are Majorana
particles and for $sin^2\theta_{13}\lesssim 10^{-5}$.}
\label{smalltheta13}
\end{center}
\end{figure}

We show in \figrefer{smalltheta13} the SRN flux for the normal mass
hierarchy (red/dotted curve) with $m_1 = 0.001 \; {\rm eV}, \; m_2 =
0.008 \; {\rm eV}, \; m_3 = 0.05 \;{ \rm eV}, \; {\rm and}\;
M_G =
173\; {\rm eV}$, the inverted mass hierarchy (blue/dashed curve)
with $m_1 = 0.05 \; {\rm eV}, \; m_2 = 0.05 \; {\rm eV}, \; m_3 =
0.008 \; {\rm eV}, \; {\rm and}\; M_G = 1225\; {\rm eV}$, and the
SRN flux without interactions (black/solid curve).

\subsection{\label{detection}Signal Detection}
\begin{figure}[h]
\begin{center}
\includegraphics[width=4in]{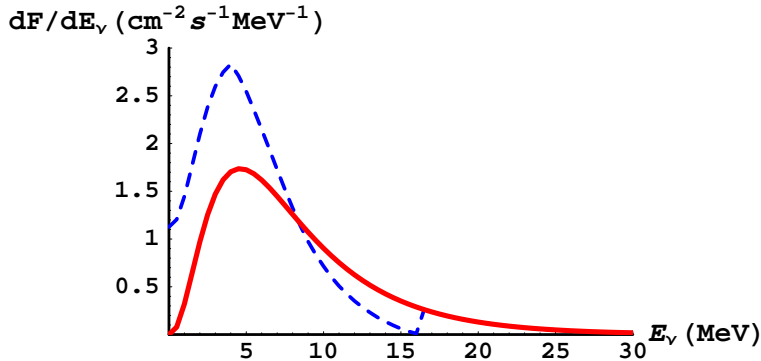}
\caption{The SRN electron antineutrino flux without interactions
(red solid curve), and with interactions and inverted mass hierarchy
(blue dashed curve) when neutrinos are Majorana particles.}
\label{16MeVplot}
\end{center}
\end{figure}

\begin{figure}[h]
\begin{center}
\includegraphics[width=4in]{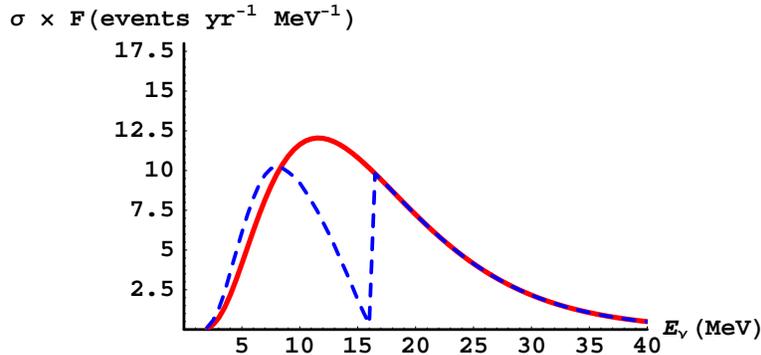}
\caption{The event rates for HyperKamiokande without interactions
(red solid curve) and with interactions (blue dashed curve) when neutrinos are
Majorana particles, for the inverted mass hierarchy.}
\label{16MeVweightedplot}
\end{center}
\end{figure}

Here we show an example of an inverted mass hierarchy.  The resonant
energy for the $m_1$ and $m_2$ neutrino mass eigenstates is taken
to be 16 MeV, which for $m_1\sim m_2\sim 0.05\,{\rm eV}$ gives
$M_G\sim 1265\,{\rm eV}$.
We present both the expected SRN flux as well as the
SRN flux folded with the cross-section given in \eqrefer{crosssection}
in \figrefer{16MeVplot} and \figrefer{16MeVweightedplot}.

Comparing the blue (dashed) curve to the red (solid) curve in
\figrefer{16MeVweightedplot} we see that there is a depletion of the SRN
event rate above approximately 8 MeV if the SRNs interact via the light
scalar at resonance with the cosmic background neutrinos (blue dashed curve).
This depletion is present up to the location of the dip, which in this
case is 16 MeV.  The main source of background, assuming the addition of Gd
to the water Cerenkov detector, are nuclear reactor electron antineutrinos,
but this background is small above neutrino energies of about 12
MeV~\cite{FogliBeacom,Hall:2006br}.  This background is dependent
on the location of the experiment and could be nearly absent.
Clearly the detector which is not near nuclear reactors would have a better chance of
seeing the signal for lower values of
the resonance energy~\cite{FogliBeacom,Hall:2006br}.

To demonstrate the significance of our signal we look at an energy of 15 MeV
(where the effect is most significant and the reactor background is negligible).  At this energy we expect approximately
11 events per year at HyperK from the SRN flux without new interactions
(red solid curve).  However, if the neutrinos interact via the light
scalar at resonance, then we instead predict approximately 2 events per
year (blue dashed curve) at this energy bin (assuming a 2 MeV bin).
The average fluctuation in the number of events expected in 5 years with
no interactions can be estimated as $\sigma \sim \sqrt{55} \sim 7.5$ events.
If we take as our signal the number of events expected without interactions
minus the number of events expected with interactions (i.e., the event deficit) over the 5 year period (this is $55 - 10 = 45$ events), then in 5 years
one expects approximately a $6\sigma\; (45/7.5)$ effect. A similar
analysis can be performed for the previous cases discussed in this paper,
but since the depletion in these cases is at lower energies one must
pay careful attention to the reactor background.

This analysis also requires a side band study, in order to determine the
SRN flux in a region where interactions are ineffective (in our present
example, this would be above 16 MeV).  This would provide the overall
normalization necessary for establishing the background.  Clearly, once
the shape of the signal without resonance and the non-SRN background are
known, a more sophisticated analysis (including a bin-by-bin fit to
the shape of the curves) can be achieved, and may even provide enhanced
significance.  This, however, is beyond the scope of this work, whose
aim we regard to be a discussion of the qualitative aspect of our new
physics signal.

For a liquid argon detector, if the resonance energy is above the
cutoff for the solar neutrino flux at about 19 MeV, the integrated
number of SRN electron neutrino events would be visibly reduced in the presence
of new interactions.  This is shown for a normal mass hierarchy and an inverted
mass hierarchy in \figrefer{liquidargon} and \figrefer{liquidargoninverted}
respectively.  While our total number of events is a conservative estimate
(we use a $z$ evolution that is flat above $z = 1$, instead of stronger
dependence~\cite{Ereditato:2005ru}), the depletion of events up to the
resonance cutoff is a robust feature.  We find that there is approximately a 25\%
reduction in the number of electron neutrino absorption events with new interactions
compared to without new interactions for both the normal and inverted mass hierarchies
for the energy window from 19 MeV to 40 MeV.

\section{\label{conclusion}Conclusion}
The late-time neutrino mass generation models could be tested by
detecting unique features of the SRN flux in both its electron
antineutrino and neutrino components (for other tests of new
physics that can be done with the SRN flux see~\cite{nudecay}). To
illustrate this new effect we have considered an Abelian U(1) model
that generates neutrino masses at low scales. However, it is clear
that the main features would still hold for a more complicated,
non-Abelian model, although these models are already more
constrained by BBN considerations. For example, one could still have
observable dips in the SRN spectrum if the resonances are in a
desirable energy window, but the couplings are no longer
proportional to the neutrino masses, and so some predictive power is
lost.  However, one could hope to correlate the observations of the
dip locations in the SRN spectrum with signals proposed to be
present in the cosmic microwave background~\cite{global} for this
case.  We expect that the future generation water Cerenkov detectors
enriched with gadolinium such as UNO, HyperKamiokande, or MEMPHYS
would be able to detect a substantial number of SRN antineutrino
events in a year~\cite{nextgenexp}.  Note that the threshold
for this is on the order of 10 MeV and depends on the location of
the detector, especially due to reactor
backgrounds~\cite{FogliBeacom, Hall:2006br}.
The effects of smearing due to the energy
resolution of the water cerenkov-type detector needs to also be taken into
account in a detailed analysis.  For a Gaussian energy resolution
function with width $\Delta$ ($\Delta/{\rm MeV}\sim0.6\sqrt{E/{\rm MeV}}$),
the smearing is only at most a few MeV~\cite{FogliBeacom} in the energy domain
we considered.  This smearing is then always smaller than the width of the depletion features considered. The neutrino component
of the SRN flux could be detected by a large 100 kton liquid argon
detector~\cite{Ereditato:2005ru}.
If there are neutrino-neutrino
interactions through the light scalars present in these models,
there is a possibility to distinguish between normal and inverted
mass hierarchies and Dirac versus Majorana neutrinos, as well as to
determine the absolute values of the neutrino masses.  The
ability to distinguish the neutrino mass hierarchy is independent of
whether supernova neutrino flavor evolution is adiabatic or
non-adiabatic.

The qualitative features of
the signal of new interations via light scalar, such as the depletion,
enhancement at lower energies, and the possibility to distinguish
between neutrino mass hierarchy, as well as the nature of the
neutrino are independent of the theoretical model for
the supernova neutrino energy spectrum, which
 predict slightly different shape and wider range of average
energies for different neutrino (antineutrino) flavors than the
KRJ model~\cite{Totani:1997vj, Keil:2002in, Thompson:2002mw}.
 This is not surprising because the produced neutrino spectrum
does not depend
on the detailed shape and normalization of the initial supernova neutrino
fluxes but rather on the coupling of the scalar to the final neutrino
mass eigenstates.  As shown in section A (3) of the first
paper~\cite{Goldberg:2005yw}, the presence of a deep dip is universal,
and its position depends only on the masses of the scalar and the
target neutrino, rather than any feature of the neutrino spectrum.
For a normal neutrino mass hierarchy there is an overall depletion of the
SRN flux, while for an inverted neutrino mass hierarchy there is an
enhancement of the SRN flux at low energies and a region of
depletion at higher energies. If a sterile neutrino with a much
larger mass than the active neutrinos were also to couple to the new
scalar, then independent of the details of the masses of the active
neutrinos the effect would be almost complete depletion of the
spectrum in some energy window since the scalar would decay
predominantly to the massive sterile neutrino.

\begin{figure}[!hb]
\begin{center}\hspace*{-.1cm}
\includegraphics[height=8.5cm, width=13.5cm]{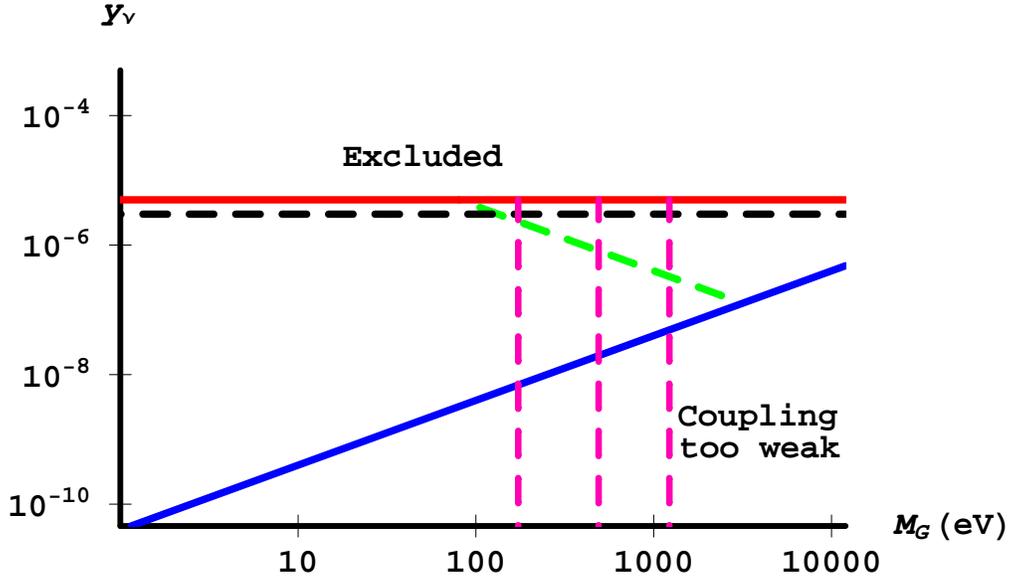}
\caption{The cosmological bounds and
the regions for the supernova neutrino spectrum distortion due to the resonance
and non-resonance processes for a single Majorana (Dirac)
neutrino for an abelian (non abelian) model
 are shown in ($y_\nu,M_G$) plane.
The region above the red (solid horizontal) line
is excluded by the BBN constraint (for the Dirac case), SN cooling (for Majorana case)
 and due to the observation of (undegraded) SN1987A neutrinos.  In the region below
the blue (solid slanting) line the mean free path is too long for
the resonance to occur. The region above
the green (dashed slanting) line, which is relevant only for
 the non abelian Majorana case, is the region excluded by the BBN constraint.
 The region above the black (dashed horizontal) line is the region of the
 future experimental sensitivity to the
 observation/non-observation of the SRN neutrinos due to  {\em non-resonant processes.}
The vertical dashed lines correspond to the minimum values of $M_G$
that lead to a depletion signal that is above the
nuclear reactor antineutrino background for the
neutrino masses of $0.001$ eV, $0.008$ eV, and $0.05$ eV from left
to right respectively. } \label{figcompar}
\end{center}
\end{figure}

All of these signals, and especially their observation, depend on
the parameters of the model.  In \figrefer{figcompar} we show
constraints on the parameter space for which the SRN effects can be
obtained in the $y_\nu - M_G$ plane. The signals proposed here are
present in the SRN flux only if the couplings of the neutrino mass
eigenstates to the scalar are larger than the condition given in
\eqrefer{lowercoupling} for a given value of $M_G$. This condition
comes from requiring that the mean free path for absorption of
a SRN neutrino on a cosmic background neutrino is much smaller than
the Hubble scale~\cite{Goldberg:2005yw}.  It is a sufficient
condition to guarantee the absorption of all three neutrino flavors.
This lower bound on the coupling is represented by
the diagonal blue (solid) line. If
the resonance energy is below 12 MeV then there is a large
background from nuclear reactor antineutrinos~\cite{FogliBeacom,Hall:2006br}.
To have a significant signal we take $M_G$ to be above
$\sqrt{2m_\nu E_{\nu,{\rm min}}^{Res}}\;$, where
$E_{\nu,{\rm min}}^{Res}$ is approximately 15 MeV.
These threshold values are represented
by the three vertical red dashed lines which are calculated for values of $m_\nu = 0.001
\; {\rm eV},\; 0.008\; {\rm eV},\; {\rm and}\; 0.05\; {\rm eV}.$  If the mass
of the scalar is larger than these values, then the signal would
be above the reactor background.  Similarly, the signal would not be
observable if the mass of the scalar is large so that the heaviest neutrino mass
eigenstate has a resonance energy in the region where the SRN flux
is small. We also show
the constraint imposed by BBN considerations, which is
similar to the bound obtained from SN cooling and to the bound
from the observation of undegraded SN1987A neutrino
flux~\cite{Goldberg:2005yw}. The SRN flux is also sensitive to the
non-resonant process, for example $2\nu\rightarrow\phi\rightarrow
2G\rightarrow 4\nu$, but only in a very small region of the
parameter space, above the horizonal black dashed line and below
the horizontal red solid line~\cite{Goldberg:2005yw}. The area
above the diagonal green dashed line corresponds to the
BBN constraint for a non-abelian Majorana case.  We note that there is
still a large range of parameter space where the couplings are large
enough to give SRN flux modification in an energy window that large
neutrino detectors could directly probe.

We have shown that the cosmic background neutrinos interacting with
supernova relic neutrinos through exchange of the light scalar lead
to significant modification of the SRN flux observed at earth. These
signals would be detectable for a large region of parameter space,
in some cases at a significance of more than 5$\sigma$,
and measurements of the presence of these effects are well within
the reach of the next-generation water Cerenkov detectors enriched
with gadolinium, or a large 100 kton liquid argon detector.
Specifically we have shown that the changes induced in the flux by
the exchange of the light scalars might allow one to distinguish
between neutrinos being Majorana or Dirac particles, the type of
neutrino mass hierarchy (normal or inverted or quasi-degenerate),
and could also possibly determine the absolute values of the
neutrino masses.  An interesting feature is that the ability to
distinguish neutrino mass hierarchy does not depend on the dynamics of
the flavor
evolution of neutrinos leaving the supernova (whether it is adiabatic or
non-adiabatic), or on the specific shape and normalization
of the initial supernova neutrino flux.  Note that the hierarchy
determination can be made by solely looking at the spectrum of
supernova relic electron antineutrinos, without need to do the measurement of
the flux of supernova relic electron neutrinos.  In addition, the
modification of the SRN flux in any of the proposed scenarios is a
clear indication of the presence of the cosmic background neutrinos
left over from the era of Big Bang Nucleosynthesis.

\noindent {\bf Acknowledgements}

We thank Tom Weiler for many discussions and for his valuable
comments and suggestions.
This research was supported in part by DOE under contracts
DE-FG02-04ER41319 (JB and IS), DE-FG02-04ER41298 (JB and IS),
DE-AC03-76SF00098 (GP) and NSF under grant PHY-0244507 (HG).  JB and
IS would like to thank LBL Theory Group for their hospitality while
this work was being completed.


\begin{thebibliography}{10}
\bibitem{instdecay}
The decay is almost instantaneous compared to a
Hubble time: the decay rate in the lab frame of the scalar produced
at resonance is easily calculated to be $\sim (y^2/16\pi)\ m_\nu,$ which
is much larger than an inverse Hubble time if $y \gg 4\times 10^{-16}
(1\ {\rm eV}/m_\nu))^{-1/2}.$ In the light of the much stronger
requirement \eqrefer{lowercoupling}, this is easily satisfied for all neutrino
masses we consider.
\end{thebibliography}
\end{document}